\documentclass[11pt]{article}
\pdfoutput=1
\usepackage{amsmath,epsf,amssymb,latexsym,enumerate,cite,shadow,array,color}
\usepackage{slashed}
\usepackage{graphicx,wasysym}

\DeclareFontFamily{U}{rsf}{} \DeclareFontShape{U}{rsf}{m}{n}{
  <5> <6> rsfs5 <7> <8> <9> rsfs7 <10-> rsfs10}{}
\DeclareMathAlphabet\Scr{U}{rsf}{m}{n} \makeatletter
\@addtoreset{equation}{section} \makeatother

\def\be{\begin{equation}}
\def\ee{\end{equation}}
\def\ba{\begin{array}}
\def\ea{\end{array}}

\newcommand{\beq}{\begin{equation}}
\newcommand{\eeq}[1]{\label{#1}\end{equation}}
\newcommand{\bea}{\begin{eqnarray}}
\newcommand{\eea}[1]{\label{#1}\end{eqnarray}}

\def\K{K{\"a}hler}
\def\vp{{\varphi}}
\usepackage{ifpdf}
\ifx\pdfoutput\undefined
   \pdffalse
\else
   \pdfoutput=1
   \pdftrue
  \usepackage[pdftex]{hyperref}
\newcommand{\rf}[1]{(\ref{#1})}

\begin{document}

\begin{titlepage}
\hfill CERN-PH-TH/2013-214\\

\hskip 1.5cm

\begin{center}

{\huge \bf{Higher Order Corrections in \\

\vskip 0.8 cm 

Minimal Supergravity Models of Inflation}
 }

\vskip 0.8cm  

{\bf \large Sergio Ferrara$^{1,2}$, Renata Kallosh$^3$,  Andrei Linde$^3$ and Massimo Porrati$^4$}  

\vskip 0.5cm

\noindent $^{1}$ Physics Department, Theory Unit, CERN, CH 1211, Geneva 23, Switzerland\\
$^{2}$ INFN - Laboratori Nazionali di Frascati, Via Enrico Fermi 40, I-00044 Frascati, Italy~\footnote{On leave of absence from Department of Physics and Astronomy, University of California Los Angeles, CA 90095-1547 USA}\\
$^3$ Department of Physics and SITP, Stanford University, Stanford, California
94305 USA\\
$^{4}$ {CCPP, Department of Physics, NYU
4 Washington Pl. New York NY 10016, USA}

\end{center}

\vskip 1 cm

\begin{abstract}
We study higher order corrections in new minimal supergravity models of a single scalar field inflation. The gauging in these models leads to a massive vector multiplet and the D-term potential for the inflaton field with a coupling $g^{2} \sim 10^{{-10}}$.   In the de-Higgsed phase with vanishing  $g^2$, the chiral and vector multiplets are non-interacting, and the potential vanishes.  We present generic manifestly supersymmetric higher order corrections for these models.  In particular, for a supersymmetric gravity model  $-R+ R^2$ we derive manifestly supersymmetric corrections corresponding to $ R^n$. The dual version corresponds to a standard supergravity model with a single scalar and a massive  vector. It includes, in addition, higher Maxwell curvature/scalar interaction terms of the Born-Infeld type and a modified D-term scalar field potential. We use the dual version of the model to argue that higher order corrections do not affect the last 60 e-foldings of inflation; for example the $\xi R^4$ correction  is irrelevant as long as $\xi< 10^{24}$.

 \end{abstract}

\vspace{24pt}
\end{titlepage}

%{\parskip -5pt \tableofcontents}

\tableofcontents

\section{Introduction}

In this paper we will analyze some of the issues related to inflationary cosmology from the perspective of a new formulation of inflationary theory in the context of supergravity with vector or tensor multiplet  proposed in~\cite{Ferrara:2013rsa,vp}. 

The first of these issues is the naturalness of the choice of the inflationary potential. The second is the issue of quantum  corrections. These problems are closely related, but not equivalent. For example, one may consider the simplest chaotic inflation potential ${m^2\over 2}\phi^2$  with $m \sim 6 \times 10^{-6}$, as required by the normalization of the amplitude of inflationary perturbations  \cite{Linde:1983gd}. If this field very weakly interacts with other fields, then quantum corrections to this potential, including quantum gravity corrections, can be extremely small. However, adding higher order terms ${\lambda_{n}}  \phi^{2n+2}$ with $\lambda_{n} = O(1)$ could strongly modify the potential at $\phi \gtrsim 1$.  

There are several different ways to protect flatness of the inflationary potential. In particular, as we will see, this problem does not appear in the context of the new approach to supergravity inflation proposed in \cite{Ferrara:2013rsa}: If we take the first term in the expansion of the potential proportional to $g^{2}$, which takes the role of the parameter $m^{2}$ in the example described above, then all terms higher order in the inflaton field will be automatically suppressed by the small constant $g^{2}$, or by even smaller factors $g^{2n}$.

We will also discuss a similar issue in the context of the inflationary model $-R+ \alpha R^{2}$ \cite{Starobinsky:1980te,Kofman:1985aw,Whitt:1984pd}.  Its supergravity implementation was given  in old minimal supergravity in \cite{Cecotti:1987sa} and in new minimal supergravity in
  \cite{cfps} and discussed recently in \cite{Farakos:2013cqa} and in \cite{Ferrara:2013rsa}. In  \cite{Farakos:2013cqa} the supersymmetric generalization   of the $\xi R^4$ was given. 
  It was also argued in \cite{Farakos:2013cqa} that the  term $\xi R^4$ typically destroys inflationary predictions of the original model. However, we will show that this conclusion was based on the implicit assumption that it is natural to consider the term $\xi R^4$ with $\xi \gtrsim 10^{24}$, because this coefficient was represented as $\xi \sim s^2/ g^6\approx  s^2 \, 10^{30}$,  with fixed $s$ and $g^{2} \sim 10^{{-10}}$.
 We will show that the dependence of $\xi$ on inverse powers of $g^{6}$ is inconsistent with the de-Higgsed phase of our dual model, where the limit to $g\rightarrow 0$ is well defined and corresponds to  non-interacting chiral and massless vector multiplets. 
  
We will focus on the  study of
new minimal supergravity models of inflation \cite{Ferrara:2013rsa}, \cite{vp} of a single scalar field inflation defined by a choice of the Jordan frame function $\Phi(C)= e^{{2\over 3} J(C)}$.  The gauging in these models with the gauge coupling $g^2$ leads to a massive vector multiplet  and a D-term potential for  a single real scalar. The generic F-term potential is absent since in the Higgsed phase with non-vanishing $g^2$ our minimal supergravity model has no chiral multiplets.  In the de-Higgsed phase in the limit of a vanishing  $g^2$ the chiral and vector multiplets are non-interacting and there is no potential. This is the trademark of our class of models:  the potential (with and without corrections)  vanishes in the limit  $g\rightarrow 0$, vectors and 2 scalars become massless, whereas at $g\neq 0$ there is a potential for a single scalar, and the 3-component  vector  is massive.

We study these models in the presence of corrections  due to terms of higher order in the field strength $W_\alpha$, which correspond to higher order
$D$ terms in the supersymmetric  action.  Each extra power of $W^2$ adds to the action extra powers of $(F^2_{\mu\nu}-D^2)$.
This $D$-dependence on higher powers of the superfield $W$ leads to a modification of the inflationary potential, which we would like to evaluate. 
We provide a gauge-invariant superconformal action for these models in the presence of higher derivative corrections and gauge-fix it. This results in the Einstein frame  supergravity  action with higher powers of $F_{\mu\nu}$ and with a corrected potential. In the case of a general choice of the model with the Jordan frame function $\Phi(C)$, our higher order terms with $W^{N}$ may depend on arbitrary functions $\Psi_N(C)$. Properties of these terms will be studied and their role during inflation will be evaluated for the general case. An important guideline for these higher order terms will come from their consistency with the de-Higgsed phase of the model at $g\rightarrow 0$ when the chiral and vectors multiplets are decoupled. The issue of higher order corrections in other cosmological models based on supergravity, such as \cite{Ellis:2013nxa}, \cite{Fre:2013dha}, has to be studied separately.

A special case in our models, with $\Phi(C)=-Ce^C$ and gauge coupling $g^2 = 1/9\alpha$, has a dual version corresponding to pure supergravity action $-R+\alpha R^2$ without scalars or vectors, corresponding to inflationary models \cite{Starobinsky:1980te,Kofman:1985aw,Whitt:1984pd}. In this case  the higher powers of $W^n$ correspond to a supersymmetric version of $R^n$, which we construct here. In this case the generic functions $\Psi_n(C)$ are identified from the requirement of duality to pure gravitational multiplet without other multiplets. 
We present specific expressions for these models and confirm our general study of higher corrections.

Before going any further, we should clarify certain terminological issues. The words ``old minimal'' and ``new minimal'' supergravity  in our manifestly superconformal actions mean that in ``old minimal'' the conformal compensator superfield is a chiral multiplet, 
``new minimal''  means that the conformal compensator superfield is a linear multiplet. We study in this paper the minimal supergravity inflation models \cite{Ferrara:2013rsa}, which means that the matter multiplet in our model in the Higgs phase is a massive self-interacting vector multiplet; we have a massive vector and a real scalar and a potential due to gauging.  In the de-Higgsed phase with $g=0$ we have a complex scalar and a massless vector which are, however, decoupled. To these models we add terms with higher powers of superfields, allowed by the symmetries of the problem.

In Section 2 we discuss the general issue of higher order terms in inflationary models. In Section 3 we study a superconformal and gauge invariant action with Einstein gravity, a single scalar and a massive vector, and introduce  higher order terms in vector multiplet superfields consistent with the symmetries of our models in the case of general $\Phi(C)$.  In Section 4 the case of supersymmetric $R^n$ corrections to $-R+\alpha R^2$ model is studied in the dual version of the theory with Einstein gravity, a single scalar and a massive vector. We  introduce  higher order terms in vector multiplet superfields consistent with the symmetries of our models in the case of $\Phi(C)=-Ce^C$, which correspond to supersymmetric $R^n$ terms. This provides a special example of our general formula found in Section 3. In Section 5 we evaluate the effect of these corrections on inflation. Finally, in Section 6 we give a summary of our results.

\section{Higher order corrections: The history of the problem}

For a long time, it seemed rather difficult to implement inflationary theory in supergravity. For example, the simplest F-term potentials contain the overall factor $e^{K}$. The popular choice of the \K\ potential $Z\bar Z$ resulted in extremely steep potentials $\sim e^{Z^{2}}$, which are not suitable for the simplest chaotic inflation models \cite{Linde:1983gd}. Recently two classes of models have been found which solve this problem and allow construction of inflationary models with nearly arbitrary choice of inflationary potential, which should allow to fit {\it any}\, set of observed values of the inflationary parameters $n_{s}$ and $r$. The first of these two classes is based on the old minimal supergravity, with inflation driven by the F-term potential and the inflaton field being part of a chiral multiplet \cite{Kallosh:2010ug}. The second class is based on new minimal supergravity with a massive vector or tensor multiplet; the evolution of the inflaton field in this class of theories is determined by the D-term potential \cite{Ferrara:2013rsa,vp,cfps,Farakos:2013cqa}. In this paper we concentrate on the latter class of models.

Another set of problems is related to naturalness of the choice of the inflationary potential, and to quantum corrections to the potential. Fore example, one may consider the simplest chaotic inflation potential 
 \be\label{standard}
 V(\phi) = {m^2\over 2}\phi^2
 \ee
with $m \sim 6 \times 10^{-6}$  \cite{Linde:1983gd}. One can suppress interaction of this fields with other fields to avoid quantum corrections to the inflationary potential, but one cannot suppress interaction of this field (or its potential) with gravity. Fortunately, quantum gravity corrections in this model are expected to be suppressed by higher powers of $V''$ and $V$, in Planck units. During the last 60 e-foldings of inflation, $V''$ and $V$ are of order $10^{-10}$, so they can be safely ignored \cite{Linde:2005ht}.

However, even if quantum corrections are negligible, one may wonder whether the initial assumption that $V(\phi) = {m^2\over 2}\phi^2$
with $m \sim 6 \times 10^{-6}$ is natural. One may try to represent $V(\phi)$ as a series in powers of $\phi/M_{p}$, where in our notation $M_{p} = 1$. This is an important assumption, because in general the scalar field itself may not have any invariant meaning. A more appropriate expansion parameter would be $g\phi$, where $g\phi$ is the typical  mass of elementary particles interacting with the field $\phi$. Let us temporarily ignore this point, which is going to be very important for us later, and write an expected expansion of the potential in terms of the field $\phi$:
\be\label{serieschaotic}
 V =  {m^2\over 2}\phi^2 +  \sum^{\infty}_{n=0} {\lambda_{n}}{\phi}^{2n+4} \ .
 \ee
In the original model ${m^2\over 2}\phi^2$ the last 60 e-foldings of inflation begin at $\phi \approx 15$. The requirement that at $\phi = O(10)$ all higher order terms are smaller than ${m^2\over 2}\phi^2$ with $m^{2} \sim 4\times 10^{-11}$ does not seem particularly natural, unless we know some reason why $\lambda_{n} 10^{2n+4}$ is smaller than $10^{-11}$, for all $n$. 

There are several ways to address this problem, involving the existence of flat directions of the potential protected by some symmetry \cite{Freese:1990rb,Silverstein:2008sg}, or some mechanisms which may flatten the potentials \cite{Salopek:1988qh,Sha-1,Ferrara:2010in,Dong:2010in,Kallosh:2013pby,Kallosh:2013hoa}. In some cases, flatness of the potential is inseparably linked to the smallness of the first term in the expansion. If so, no additional $O(10^{-10})$ or $O(10^{-11})$ fine-tuning of the higher order terms in $\phi$ is required. For example, in the F-term models of Ref. \cite{Kallosh:2010ug} one can interpret the flatness of the potential as a consequence of the smallness of the constant $\lambda$ in the superpotential $\lambda S f(\Phi)$, which leads to the potential $\sim \lambda^{2} f^{2}(\phi)$. Here $S$ is the sgoldstino field, which vanishes during inflation, and $\phi$ is the flat direction of the potential, which may correspond to either the real or the imaginary part of $Z$, depending on the choice of the \K\ potential. 

A similar interpretation emerges even more naturally in the new approach to inflation in supergravity \cite{Ferrara:2013rsa}. We will see that the inflaton potential in this class of theories has a structure which can be schematically represented as
\be
V = {g^{2}\over 2} F(\phi) + O(g^{4}) \ ,
\label{A}\ee 
where $g^{2}$ can be arbitrarily small, e.g. $g^{2}\sim 10^{-10}$. Note that this small parameter stays in front of the function $F(\phi)$, so that the whole potential disappears in the limit $g^{2}\to 0$. This makes a lot of difference as compared with (\ref{serieschaotic}), where only the first term in the series was small, and each new term required individual fine-tuning to make it small. The choice of the small parameter $g^{2}$ is a legitimate price to pay for the description of the anomalously small amplitude of perturbations of metric in the observable part of the universe, just like the small coupling of the electron to the Higgs field $\sim 2 \times 10^{{-6}}$ is a legitimate price to pay to account for its anomalously small mass as compared to the mass of the top quark. But we pay this price only once; the only extra requirement is a choice of a function $F(\phi)$ which should be sufficiently smooth in a limited interval of the values of the inflaton field and should account for two other experimentally measurable parameters, $n_{s}$ and $r$. This is similar to what we routinely do in particle physics: We find particular values of the parameters of a theory which are required to describe a number of observationally measured quantities.

Another example which we may consider here is  the model $-R+R^{2}$ \cite{Starobinsky:1980te}: 
\be\label{star}
L= { \sqrt{-g}}  f(R) = { \sqrt{-g}} \left(-{1\over 2} R+{R^2\over 12M^2}\right) \, .
\ee
Here $M \sim 1.3\times 10^{-5}$ is the `scalaron' (inflaton) mass, in Planck units. Thus the term $R^{2}$ added to the Einstein gravity has a very large coefficient $\sim 10^{9}$. The equation for the Hubble constant during inflation in this model implies that the last 60 e-foldings of inflation in this model begin when  $R = 12 H^{2}  \sim 240 M^{2}$ \cite{Kofman:1985aw}. Not surprisingly, it happens at the time when the first and the second term in (\ref{star}) are of the same order.  More generally, the value of $R$ in this scenario is equal to $4M^{2} N$ at the time $N$ e-foldings prior to the end of inflation.
 
 Now let us check what may happen if we modify this model by adding to it higher order terms $R^{N}$. As we will see later (see also \cite{Farakos:2013cqa}) the lowest term of this type compatible with supersymmetry is $R^{4}$: 
\be\label{star2}
L= { \sqrt{-g}}  f(R) = { \sqrt{-g}} \left(-{1\over 2} R+{R^2\over 12M^2} + \xi R^{4}\right) \, .
\ee
Using the estimate $R = 4M^{2} N$ at the beginning of the last $N$ e-foldings of inflation, one finds that the term $\xi R^{4}$ is smaller than ${R^2\over 12M^2}$ in this expression unless  $\xi \gtrsim 10^{{-2}} M^{-6} N^{{-2}}$. In particular, for $\xi = O(1)$ the term $R^{4}$ is negligible during the last $10^{14}$ e-folds of inflation. The term $R^{4}$ becomes of the same order as the term ${R^2\over 12M^2}$ during the last 60 e-foldings of inflation only if $\xi \gtrsim  10^{25}$. This does not seem particularly natural, so one could conclude that the model (\ref{star}) is very stable with respect to higher order corrections. As we will see, a detailed investigation of this issue in the context of \cite{Ferrara:2013rsa} confirms this conclusion.

However, on the basis of a very similar investigation, the authors of   \cite{Farakos:2013cqa} came to an opposite conclusion. They finished their paper by the statement that this model ``suffers from the same difficulty one encounters when trying to embed a model of inflation in supersymmetry where the flatness of the potential is easily destroyed by supergravity corrections.'' One of the goals of our investigation is to examine this problem, which should help us to take a fresh look at the whole issue of higher order corrections in various models of supergravity inflation.

\section{Superconformal action with   higher order corrections for generic $\Phi(C)$ models}
In \cite{Ferrara:2013rsa} we used as a starting point the following gauge invariant superconformal action
\beq
{\cal L}= -S_0 \bar{S}_0 \Phi(\Lambda+\bar{\Lambda} + g V)_D + {1\over 2 } [W_\alpha(V) W^\alpha(V) + h.c]_F .
\eeq{m1}
The action has  superconformal symmetry and at $g\neq 0$ it has a St\"uckelber-type gauge symmetry 
\beq
\Lambda \rightarrow \Lambda + \Sigma, \qquad V \rightarrow V -{1\over g} (\Sigma +\bar{\Sigma}),
\eeq{m2}
where $\Sigma$ is a chiral superfield. The second terms in the action is gauge-invariant under the shift of the vector by a chiral plus anti-chiral field as shown in \rf{m2}.
Here
 $\Phi$ is an arbitrary positive function of its argument
\beq
\Phi = e^{{2\over 3} J} \, , \qquad    \,  J\equiv  {3\over 2} \log \Phi  .
\eeq{m3}
At $g\neq 0$ the chiral/anti-chiral superfields $\Lambda, \bar \Lambda$ in \rf{m1} are absorbed into a vector multiplet $gV$, which corresponds to gauge-fixing the gauge symmetry \rf{m2}. The classical part of the bosonic component action is 
\beq
{\cal L}= -{1\over 2}R +{1\over 2} J'' (C)\partial_\mu C\partial^\mu C -{1\over 4} F_{\mu\nu}(B)F^{\mu\nu}(B) 
 +{1\over 2} D^2  + g J'(C) D+ {g^2\over 2} J'' B_\mu B^\mu.
\eeq{m4}
When the auxiliary field $D$ is evaluated on shell using its equations of motion, we find
\beq
{\cal L}= -{1\over 2}R +{1\over 2} J'' (C)\partial_\mu C\partial^\mu C -{1\over 4} F_{\mu\nu}(B)F^{\mu\nu}(B) + {g^2\over 2} J'' B_\mu B^\mu
 -{g^2\over 2} J'^2(C).
\eeq{m5}
Note that model at $g=1$ is an example of self-interacting massive vector multiplet given in \cite{vp} and it also belongs to the general class of  supergravity models studied in \cite{cfps,cfgvp}. 

The essence of the de-Higgsed phase here is that in components $B_\mu$ has absorbed the term ${1\over g} \partial_\mu a$ which would replace the mass of the vector term ${g^2\over 2} J'' B_\mu B^\mu$ by the kinetic term of the axion ${1\over 2} J'' (\partial_\mu a)^2$ in the limit $g\rightarrow 0$.

We may present the same model with the gauge-symmetry fixed  in the form
\beq
{\cal L}= -S_0 \bar{S}_0 \Phi(U)_D + {1\over 2 g^2} [W_\alpha(U) W^\alpha(U) + h.c]_F \ , 
\eeq{m6}
where $gV=U$. In components, the bosonic Lagrangian is
\beq
{\cal L}= -{1\over 2}R +{1\over 2} J'' (C)\partial_\mu C\partial^\mu C -{1\over 4 g^2}  F_{\mu\nu}(\tilde B)F^{\mu\nu}(\tilde B) 
 +{1\over 2 g^2} \tilde D^2  +  J'(C) \tilde D+ {1\over 2} J'' \tilde B_\mu \tilde B^\mu.
\eeq{m7}
When the auxiliary field $\tilde D$ is set on shell, we find the same potential, $V= {g^2\over 2} J'^2(C)$, but the Maxwell part of the action is not canonical. It is easy to change back to $B_\mu$ from $\tilde B_\mu$.

The reason for our discussion of this subtlety before introducing terms of higher power in superfields is the high sensitivity of the dependence on $g^2$ of higher order corrections.

To derive the Einstein frame supergravity model we gauge-fix the extra Weyl symmetry by making the floowing choice in units $M_P=1$
\be
S_0 \bar{S}_0 \Phi (U)_D=1 \ .
\label{SC}\ee

\subsection{A quartic in superfields $W$ correction}
We now add higher order terms, quartic in the $W$ superfield, so that
\bea
{\cal L}^\xi= -S_0 \bar{S}_0 \Phi(U)|_D + {1\over 2 g^2} [W_\alpha(U) W^\alpha(U) + h.c]_F
+ \xi \,  \, {W_\alpha(U) W^\alpha(U) \bar W_{\dot \alpha} (U) \bar W^{\dot \alpha} (U) \over \left(\Phi(U)S_0 \bar{S}_0\right)^{2}}~\Psi(U)|_D \, .
\eea{m8}

Here the function $\Psi(U)$ is an arbitrary function which has  zero Weyl weight.
If we were to rescale $W $ into $g W$, which corresponds to rescaling $F, D$ into $gF, gD$ so that these fields have a canonical kinetic term, we would get
\bea
{\cal L}^\xi= -S_0 \bar{S}_0 \Phi(U)|_D + {1\over 2 } [W_\alpha W^\alpha+ h.c]_F  +
 \xi \, g^4 \,  \, {W_\alpha^2  \, \bar W_{\dot \alpha}^2  \over \left(\Phi(U)S_0 \bar{S}_0\right)^{2}}~\Psi(U)|_D \ .
\eea{m9}
For  the de-Higgsed phase,   we can take the limit  $g\rightarrow 0$ with $U= \Lambda + \bar \Lambda +gV$ and $U \rightarrow  \Lambda + \bar \Lambda $
and we  find
\bea
{\cal L}^\xi _{g\rightarrow 0} \Rightarrow  -S_0 \bar{S}_0 \Phi(\Lambda+\bar{\Lambda} )|_D + {1\over 2 } [W_\alpha(V) W^\alpha(V) + h.c]_F  \ .
\eea{m10}
This is a decoupled massless vector multiplet and a chiral multiplet.

From now on we keep $g\neq 0$.  We still have to gauge-fix the superconformal symmetry in the action including corrections. To get to the  Einstein frame we need to keep a gauge-fixing \rf{SC} and the resulting higher order term becomes
\be
\xi \, g^4  \, W_\alpha^2  \bar W_{\dot \alpha}^2 \Psi( U)|_D \qquad \Rightarrow \qquad \xi \, g^4  \, D^4 \Psi( C) \ ,
\label{4}
\ee
where in general $\Psi(U)$  is arbitrary.  

The modified component action corresponding to our additional superfield action is given as follows.
The bosonic part of the component action together with higher order term \rf{4} is
\bea
{\cal L}^\xi&=& -{1\over 2}R +{1\over 2} J'' (C)\partial_\mu C\partial^\mu C -{1\over 4} F_{\mu\nu}(B)F^{\mu\nu}(B) + {g^2\over 2} J'' B_\mu B^\mu
 +{1\over 2} D^2  + g J'(C) D +\nonumber\\
 \nonumber\\
&&+\xi \, g^{4} \, [(F^{+2} -D^2)] [(F^{-2} -D^2)] \Psi(C) \ .\eea{m11}
At $g=0$ we have as before a decoupled scalar and vector multiplet so that the new terms vanish. For $g\neq 0$ the $D$ part is
\beq
{\cal L}^\xi (D) = 
{1\over 2} D^2  + g J'(C) D +\xi \, g^4 \Psi(C) D^4 \ .
\eeq{m12}
The equation for $D$ becomes
\be
\hat D(C) = -g J'(C) -4\xi g^4  \Psi(gC) \hat D^3 \ ,
\ee
and the modified $C$-dependent potential is given by 
\be
V^\xi (C)= {\cal L}^\xi (\hat D(C))= {g^2\over 2}[ J'(C)]^2+...
\ee

\subsection{ Higher order corrections}
A generic superconformal action containing higher powers of the superfield $W$ can be given in the following form:
\be
a_{k,l,p}  \, g^n  \, {W^{2} \bar W ^2 
\over \left(\Phi(U)S_0 \bar{S}_0\right)^{2}} 
\left (\Sigma {\bar W^{2}
\over \left(\Phi(U )S_0 \bar{S}_0\right)^{2}}\right )^k \left (\bar \Sigma { W^{2}
\over \left(\Phi(U )S_0 \bar{S}_0\right)^{2}}\right )^l
\left ({  ( D^\alpha W_\alpha )
\over \left(\Phi(U )S_0 \bar{S}_0\right)^{2}}\right )^p~ \Psi_{n}  ( U)|_D \ ,
\label{corr}\ee
where 
\be
n= 4+2k +2l+p\, , \qquad n\geq 4 \ .
\ee 
When the superconformal symmetry is gauge-fixed, we find
\be
a_{k,l,p}  \, g^n  \, W^{2} \bar W ^2  (\Sigma \bar W^{2})^{k} (\bar \Sigma W ^2)^{l}  ( D^\alpha W_\alpha )^p ~ \Psi_{n}  ( U)|_D \ ,
\ee
 In components, the following bosonic parts of the action are added
\beq
{\cal L}^{k,l,p}= a_{k,l,p} \, g^{n} \, [(F^{+2} -D^2)]^{1+k} [(F^{-2} -D^2)]^{1+l}   D^p \Psi_{n}  ( C).
\eeq{sm13}
The Maxwell curvature terms have a structure of the Born-Infeld type; namely, there are increasing powers of $F_{\mu\nu}$.
The $D$ part of the action has additional terms 
\be
{\cal L}^{\xi_n}(D) = {1\over 2} D^2  + g J'(C) D + \sum_n \xi_{n}   \, g^n \Psi_{n} (C) D^{n} \ .
\label{caseN}\ee
where $\xi_n$ is a sum of all possible combinations of   $a_{k,l,p}$ with $ 4+2k +2l+p=n$. The algebraic equation for $D(C)$ can be solved to produce a potential
\be
V^{\xi_n} (C)= {\cal L}^{\xi_n} (\hat D(C))= {g^2\over 2}[ J'(C)]^2+...
\ee

\section{Supersymmetric $ F(R)= -{1\over 2} R+{\alpha\over 2} R^2 + \xi_n R^n$ model and its dual with the Jordan function $\Phi(C)=-Ce^C$}

A special case of our models \cite{Ferrara:2013rsa}, when $\Phi(C)=-Ce^C$,has a dual version corresponding to pure supergravity action with bosonic part $ {1\over 2} (-R+\alpha R^2)$ without scalars and vectors,  as was shown in \cite{cfps} and studied in \cite{Ferrara:2013rsa} in the context of inflationary models \cite{Starobinsky:1980te,Kofman:1985aw,Whitt:1984pd}. The higher powers of the Maxwell superfield, $W^N$, are dual to a supersymmetric version of $R^N$ and will be constructed here. In this case the generic functions $\Psi_N(C)$ will be identified from the requirement of duality to pure gravitational multiplet.

The analysis of physical states of the supersymmetric $ -R+\alpha  R^2$ model was performed in \cite{Ferrara:1978rk} using old minimal auxiliary fields, where it was found that there is one chiral multiplet with  two propagating modes
$S, P$, which acquire a square mass ${1\over 6\alpha}$. Another chiral multiplet with the same mass and 2 propagating modes originate from a scalar mode in $ {1\over 2} (-R+\alpha R^2)$ and from the $\partial_\mu A^\mu$ auxiliary supergravity field. In the limit $\alpha \rightarrow 0$ all 4 modes become non-propagating because their mass becomes infinite. This limit corresponds to the case  where $P, S, A_\mu$ are non-propagating auxiliary fields of supergravity without the $R^2$ term,  and also the extra scalar mode disappears. The corresponding supersymmetric version of the $ -R+\alpha  R^2$ gravity was constructed in \cite{Cecotti:1987sa} and indeed, the dual model is a standard supergravity, where the $R^2$ term is traded for two chiral multiplets. The supersymmetric version of $ -R+\alpha  R^2$ in the new minimal supergravity  was constructed in \cite{cfps}. In addition to the standard supergravity there is a single scalar and 3 components of the vector field, so again 4 physical degrees of freedom replace the $R^2$ term in a supersymmetric theory.

The purpose of this paper is to present a manifestly supersymmetric version of the following pure gravity model\footnote {We will discuss the ``F(R)'' supergravity model of \cite{Ketov:2009sq} separately.  Here we just mention that it does not include a supersymmetric version of the $-R+R^2$ model, whose dual requires 4 propagating modes in addition to the gravitational multiplet.}
\beq
F(R) \equiv  -{1\over 2} R+ f(R) \, , \qquad   f(R)={\alpha\over 2} R^2 +\sum_{n\geq 4} \xi_n R^n \ .
\eeq{sm1'}
There is no supersymmetric version of the theory generalizing $-R+\alpha R^2$, which at the bosonic level includes $R^3$. However,   all powers of $R$, even and odd, starting from $R^4$ with arbitrary coefficients have a supersymmetric generalization, as we will find here.

In the new minimal supergravity there exists a superfield $W_\alpha(V_R)$ such that its second component depends on the scalar curvature $R(x)$ 
\beq
W_\alpha(V_R)= ... +\theta_\alpha R(x)+...
\eeq{sm2'}
The superconformal action depends also on a compensator $L_0$ which is gauge-fixed to $L_0=1$. Therefore, the superfield action which we will present below, which depends on the $W_\alpha(V_R), \bar W_{\dot \alpha}(V_R)$ superfields polynomially, will contain a bosonic part of the kind given in $f(R)$ in equation \rf{sm1'}. 
This will be a generalization of the manifestly supersymmetric version of the model $- R +\alpha R^2$ which was constructed  in \cite{cfps} and discussed recently in \cite{Ferrara:2013rsa} as a part of the general class of models derived  in \cite{vp}. The supersymmetric version of this model including $\xi R^4$ terms generalizing \cite{cfps} was given in \cite{Farakos:2013cqa}.

The dual version of the supersymmetric $F(R)=-{1\over 2} R+ {\alpha\over 2}  R^2 +\sum_{n\geq 4} \xi_n R^n
$ gravity is a model of a self-interacting massive vector multiplet with interaction that depends on higher powers of the vector field strength, but not on its higher derivatives. It  is easy to describe it using a chiral superconformal compensator: the higher derivative terms 
 depend on the superfields $W_\alpha(V)$ and $L_0= S_0 \bar S_0 e^V$, where
\beq
W_\alpha(V) = ... +\theta^\beta (F_{\alpha \beta}(x) +\epsilon_{\alpha \beta} D)+... \ .
\eeq{sm2'a}
Here $F_{\alpha \beta}$ is  field strength of the classically auxiliary vector, which, however, becomes propagating once the $\alpha R^2$ term is added to $R$ in the dual model, and $D$ is the auxiliary field of the vector multiplet. The superconformal gauge-fixing here is $S_0 \bar S_0 \Phi(V)=1$, where the first component of the vector superfield is a scalar $C$. The gauge coupling 
$ g^2$ of the dual model equals to the factor  ${1\over 9 \alpha}$ of the $ -{1\over 2} R+ {\alpha\over 2}  R^2$ model.

\subsection{The bosonic $-{1\over 2}  R+f(R)$ theory}
The most convenient way to write this theory is in terms of two auxiliary scalar fields, 
$\sigma$ and $\Lambda$:
\beq
e \, {\cal L}=-{1\over 2} e\, R -{1\over 2} e\,  \sigma(R-\Lambda) + e\, f(\Lambda) \ .
\eeq{sm1}
We can solve for $\sigma$ and find that $R=\Lambda$, which gives back the ``pure gravity''  action ${e\over 2} R+ e f(R)$. Alternatively,  we can rewrite the action in the form
\beq
e {\cal L}= -{1\over 2} e\, (1+\sigma)R - {1\over 2} e\, \sigma  \Lambda  + e\, f(\Lambda)  \  .
\eeq{sm2}
We may now use the Weyl rescaling identity
\be
e' R(e') (1+\sigma) = e\Big (R(e) + {3\over 2} (\partial_\mu \log (1+\sigma))^2 \Big )\, ,\qquad e'_{a\mu}= e_{a\mu} (1+\sigma)^{-1/2} \ ,
\ee
so that
\beq
 {\cal L}(e) = -{1\over 2} R -  {3\over 4}   (\partial_\mu  \log (1+\sigma))^2      +  {f(\Lambda) + {1\over 2} \Lambda \sigma \over (1+\sigma)^2} \ .
 \eeq{sm2a} 
 Let us further define
 \be
 \Lambda = DC\, , \qquad 1+\sigma=-C   \, , \qquad   \sigma/(1+\sigma) = (1+C^{-1}) \ ,
 \ee
 which brings us to
\beq
{\cal L}= -{1\over 2} R -  {3\over 4}   (\partial_\mu  \log C)^2  - {1\over 2}  { D (1+C^{-1})} +   { f  (  D C   )\over C^{2}} \ .
\eeq{sm5}

Here we may switch to the canonically normalized scalar $\vp$, where $C=-e^{\sqrt{2/3}\vp}$.
This corresponds to the scalar-dependent model which is dual to  the one defined by the generic  pure gravity action $-{1\over 2} R+f(R)$. There is one propagating scalar $C$ and one auxiliary non-propagating scalar $D$, which can be eliminated using the algebraic equation
\beq
 2 {\partial f(x)\over \partial x}\Big |_{x=DC} = C+1 \ .
\eeq{sm6}
The result is the usual Einstein gravity with a single propagating scalar $C$ and a potential where $D$ is a certain function of $C$ defined by eq. \rf{sm6}:
\be
V=    {1\over 2}  { D (1+C^{-1})} -   { f  ( D C   )\over C^{2}} \ .
\label{Pot}\ee 
In the case $f(x)= {\alpha\over 2} x^2+ \sum_{4} \xi_n x^n$  the equation for $D$ is
\beq
   \hat D  =  {1+C^{-1} \over 2  \alpha}+ \sum_{n=4} {n\over  \alpha} \xi_n \hat D^{n-1} C^{n-2}   \ .
\eeq{sm6'}
Below we will establish that the dual model with only $R^2$ present corresponds to a scalar model in \cite{Ferrara:2013rsa} with $1/\alpha = 9 g^2$. We will further obtain an equation for  $D$ as an expansion in powers of the coupling $g^2$.  
The  potential will be given by \rf{Pot} with $D$ replaced by $\hat D$.

For $g=0$, $\alpha \rightarrow \infty$,  the auxiliary field $\hat D$ vanishes and there is no potential. We will find  close resemblance to this picture in a supersymmetric version of $F(R)$ gravity. The auxiliary field $D$ of the gauged supergravity relevant to this case will be identified, up to a factor, with the auxiliary field $D$ of the Maxwell supermultiplet
$(\lambda, F_{\mu\nu}, D)$. It is not accidental therefore that in the limit when there is no gauging, there is no $D$-term potential. However, in gauged supergravity with $g\neq 0$, the auxiliary field $\hat D$ 
defines the $D$-term potential. We see this phenomenon already at the non-supersymmetric level.

\subsubsection{Example of $f(R)= {\alpha\over 2} R^2+ \xi R^4$}

We take   $f(R)= {\alpha\over 2} R^2$. We get in this case
\beq
V=   {1\over 2}{ D (1+C^{-1})}  -{\alpha  D^2 \over 2}   \  .
\eeq{sm7a}
It is useful to switch to a canonically normalized $ D_c=\sqrt \alpha D$ so that 
\beq
V= -{  D_c^2 \over 2}  +  {1\over 2\sqrt \alpha }{ D_c (1+C^{-1})}  \quad \Rightarrow \quad D_c= {1\over 2\sqrt \alpha }{  (1+C^{-1})}  \quad \Rightarrow \quad V= {1\over 8\alpha} (1+C^{-1})^2 \ .
\eeq{sm7b}
This agrees with a particular example of the model in \cite{Ferrara:2013rsa} with $V= {9\over 8} g^2 (1+C^{-1})$  after the identification
\beq
 g^2= {1\over  9 \alpha} \ .
\eeq{sm8b}
This means that we have identified the pure gravity model with $R$ and $R^2$ couplings with  the dual model described in \cite{Ferrara:2013rsa} where
 \beq
{\cal L}= -{1\over 2} R - {3\over 4}   {\partial C \partial C\over C^{2}} -  { 9 \over 8} g^2(1+C^{-1})^2 \quad \Rightarrow \quad {\cal L}_{dual}= {1\over 2} \Big (-R+{1\over 9 g^2} R^2\Big ) \ .
\eeq{sm7c}
In the canonical scalar variable $\vp$ such that $C=-e^{\sqrt{2/3}\vp}$,  the action becomes the scalar dual of the Starobinsky model \cite{Starobinsky:1980te} in the form
\beq
{\cal L}= -{1\over 2} R - {1\over 2}   (\partial \vp)^2 - { 9 \over 8} g^2 (1-  e^{-\sqrt{2/3}\vp})^2 \ .
\eeq{sm9'}
During inflation $V\sim g^2\sim 10^{-10}$.

If we  specialize to the case $f(R)= {\alpha\over 2}  R^2 + \xi R^4$, the potential term in \rf{Pot}   for $\alpha= 1/9g^2$ becomes
\be
V=   {1\over 2}  { \hat D (1+C^{-1})} -   {1\over 18 g^2} \hat D^2- \xi \hat D^4 C^2   \ .
\ee 
This is the same potential obtained in~\cite{Farakos:2013cqa} from the following supersymmetric Lagrangian, up to some difference  in notation
\beq
{\cal L}= S_0\bar{S}_0 V\exp(V)+ {1\over 2g^2} [W^2(V) + h.c.]_F + {\beta\over 4}[W^2(V) \bar{W}^2(V)/(S_0\bar{S}_0)^2 \exp(2V)]_D .
\eeq{sm7}
The reason is that the Einstein frame in old minimal supergravity is defined by $S_0\bar{S}_0 V\exp(V)=1$, so that $W^2(V) \bar{W}^2(V)/(S_0\bar{S}_0)^2 \exp(2V)=W^2(V) \bar{W}^2(V) V^2$.

\subsection{Supersymmetric $-{1\over 2} R + f(R)$ in new minimal supergravity}

The key to writing the supersymmetric completion of the $f(R)$ term in general is the use of the chiral projector $\Sigma$ \cite{Kugo:1983mv}. It maps a conformal superfield of (Weyl, chiral) weights  $(w, n=w - 2)$ into a  chiral multiplet of weights $(w+1, n= w + 1)$; therefore, $\Sigma$ has weights $(1,3)$. 

The building blocks for such a supersymmetric action include the following superfields:
$\bar{W}^2$, $L_0$  are $(3,-3)$ and $(2,0)$, respectively, 
$\bar{W}^2/L_0^2$ has weights $(-1,-3)$ so that 
$\Sigma {\bar{W}^2 \over L_0^2}$  has weights $(0,0)$.
Note that these terms will only produce the even powers of $R$. To get the odd powers one has to use the following real
superfield, which has weights $ (0,0)$.
\be
{D^\alpha W_\alpha\over L_0} = {\bar D^{\bar \alpha} \bar  W_{\dot \alpha}\over L_0}  \  .
\ee
With $\Sigma (\bar{W}^2 / L_0^2)$, $W^2$ and $h.c.$, and ${D^\alpha W_\alpha\over L_0}$ one can construct a new-minimal action  
\bea
{\cal L} &= &L_0 V_R|_D + {1\over 2g^2} [W^2(V_R) + h.c.]_F  \nonumber \\ 
\nonumber\\
&&+ \Big [\sum_{k,l,p} a_{k,l,p} {W^2(V_R) \bar{W}^2(V_R)\over L_0^2} \left(\bar{\Sigma} {{W}^2(V_R) \over L_0^2}\right)^l \left(\Sigma {\bar{W}^2 (V_R)\over L_0^2}\right)^k  \Big ( {D^\alpha W_\alpha\over L_0} \Big )^p \Big ]_D  \ .
\eea{sm9}
Here
\be
\qquad V_R\equiv \log (L_0/S_0\bar{S}_0) \ .
\ee
The  bosonic part of this  supersymmetric action is
\beq
{\cal L}_B= -{1\over 2} R +{1\over 18 g^2} R^2 + \sum_{k,l,p} a_{k,l,p}R^{4+2(k+l)+p}.
\eeq{sm10}
The old minimal formulation of theory~(\ref{sm9})  is obtained by writing it in terms of a  vector $V$ and adding the constraint $[L(U-V_R)]_D$, where $L$ is a linear multiplet~\cite{Farakos:2013cqa}.
The equation of motion for $L$ sets $U= V_R + Y + \bar{Y}$,  where   $Y$ is a  chiral field. Note that the $U$ field corresponds to the one we are using in the action \rf{m1} where the Maxwell multiplet is not canonical but takes the form $-{1\over 4g^2 } W^2$.

If the same equation is used to solve for $L_0$ in terms of $U$, $S_0, \bar{S}_0$, the fields  $Y, \bar{Y}$ become implicit functions of $U$ and $S_0,\bar{S}_0$. Their form is irrelevant because they can be reabsorbed into a redefinition of the chiral compensator: $S_0\rightarrow S_0'=S_0\exp(-Y)$, that casts $L_0$ into the form
$L_0=S_0 \bar{S}_0 \exp(U)$.
Substituting in eq.~(\ref{sm9}) and choosing the Einstein frame $S_0\bar{S}_0 U\exp U=1$ to eliminate the compensator, one obtains a Lagrangian where each $W^2$
or $\bar{W}^2$ term is multiplied by $U^2$. 

A generic term $W^2 \bar{W}^2 U^2 (\bar\Sigma W^2 )^k (\Sigma \bar{W}^2)^l U^{2k+2l} (D^\alpha W_\alpha)^p U^p $ gives, in components, a bosonic term $(F^{+2} -D^2)^{1+k}(F^{-2} -D^2)^{1+l} C^{2+2k+2l}(DC)^p$.
The canonically normalized Maxwell superfield has the action $-{1\over 4 } W^2$, so the canonically normalized bosonic action is obtained by substituting $F^\pm \rightarrow F^{'\pm}=gF^\pm $, $D\rightarrow D'=gD$, $C\rightarrow C'=C$, which gives for the case of a Maxwell action with $-{1\over 4 } (F^2-D^2)$ the following higher order terms:
\beq
{\cal L}^{can}_{k,l,p}= \alpha_{k,l,p} g^{2+p} [(F^{+2} -D^2)]^{1+k} [(F^{-2} -D^2)]^{1+l} (gC)^{2+2k+2l}  (DC)^p.
\eeq{sm12}
The contribution to the potential at $F_{\mu\nu}=0$ comes from the terms of the form 
\beq
{\cal L}_{n} = \xi_{n} \, g^{n} \, D^{n} C^{n-2} , \qquad 4+2k +2l +p= n.
\eeq{sm13a}
For example the 1st correction adds to the potential a term
\beq
{\cal L}_{4} \sim \xi_4 \, g^{4} \, D^{4} C^{2} ,
\eeq{sm14}
where $\xi_4 = a_{0,0,0}$. If we compare this expression with the generic one in \rf{4} we find that for this particular model $\Psi_4(C)= C^2$. At the level $n$ we compare with \rf{caseN} and find that  for this model
\be
\Psi_n(C)= C^{n-2} \ .
\ee
Now we are ready to evaluate our higher order corrections and find whether they may affect inflation.

\section{Higher order corrections and inflation}

We have presented  above possible higher order corrections in minimal supergravity models of inflation \cite{Ferrara:2013rsa} corresponding to higher power of superfields. Before we proceed with the analysis of these corrections we would like to analyze possible anomalies in these models associated with a chiral $U(1)$ symmetry.

\subsection{Anomalies and anomaly-like corrections}
Our models contain a pseudovector that may couple anomalously to matter. As long as the vector is massive, anomalies can be canceled by the 4D Green-Schwarz mechanism. On the other hand, if the vector mass is too small, the cutoff induced by the Green-Schwarz term \cite{Green:1984sg} (and its supersymmetric completion) can fall below the inflation scale $H={ O}(g)$. This effect persists even when the only superfield transforming under the gauge symmetry is the St\"uckelber chiral multiplet 
$\Lambda$. In this case the gauge transformation translates $\Lambda$ so all fermions are neutral. Yet they couple to the gauge fields through non-minimal interactions constructed out of the covariant derivative $D_\mu a=\partial_\mu a +g A_\mu$ according to the general supergravity formulae~\cite{cfgvp}. 
Specifically, one finds that according to \cite{vp}
\beq
{\cal L}={i\over 4}\{J'(C)\bar{\Lambda}\gamma_5 \gamma_\mu \Lambda + J'(C)\bar{\psi}_\lambda \gamma_\nu \psi_\rho \epsilon^{\mu\lambda\nu\rho} -[J'''(C)/J''(C) + J'(C)]\bar{Z}\gamma_5 \gamma_\mu Z\}(\partial_\mu a +gA_\mu)+...
\eeq{anom1}

These terms are identical to minimal coupling with an effective charge depending on 
the field $C$. The effective action $\Gamma$ is always gauge invariant, so it is a 
function of $C$, and $\partial_\mu a +gA_\mu$. To compute it directly for 
arbitrary $C$ is a massive task; luckily, we are mostly interested in the effect radiative 
corrections may give 
at $|C| \gg 1$. In this limit, the function $J'(C)$ in some of our  models tends to a 
constant ${O}(1)$, equal to 3/2 in the $-R+\alpha R^2$ supergravity case 
$J(C)= (3/2)[C +\log(-C)]$, so the couplings in eq.~\rf{anom1} become identical to 
minimal couplings with charges ${ O}(1)$ and an ``effective" gauge field $A_\mu + g^{-1}\partial_\mu a$. For constant charges the
effect of anomalies is to change the effective action by a term that, in the gauge $\partial_\mu A^\mu=0$ is completely summarized by the Green-Schwarz term
\beq
{\cal L}_{GS}=\kappa g^2 [ \Lambda W_\alpha W^\alpha  + h.c.]_F.
\eeq{anom2}
The coefficient $\kappa$ is the well known anomaly coefficient equal to $q^3/16\pi^2$ for a spin 1/2 fermion of charge $q$ and to $3q^3/16\pi^2$ for a spin 3/2 fermion of the same charge~\cite{nielsen}. In the $-R+\alpha R^2$  case, for instance, the charges are, with obvious notations, $q_\Lambda= -q_Z=q_\psi=3/2$. Therefore  the potential anomaly is only due to the gravitino.

By expanding eq.~\rf{anom2} we obtain a piece containing D-terms:
\beq
{\cal L}_{GS}= {1\over 2}\kappa g^2 C D^2 +...
\eeq{anom3}
It corrects the D-term equation in~\rf{m4} as follows
\beq
gJ'(C)+ (1+\kappa g^2 C)D=0.
\eeq{anom4}
Since $\kappa$ is ${ O}(1/16\pi^2)$, this correction is small as long as $|C| \ll 16\pi^2/g^2 \sim 16\pi^2 10^{10}$. We may estimate $C$ during inflation: in the $-R+1/3g^2 R^2$ model $|C| \sim 80$. Clearly the effect of the anomaly is negligible in these models as well as in analogous  cases.

One may also worry about another effect. The bosonic part of term~\rf{anom2} contains the standard axion-gauge field coupling which cancels the anomalous gauge transformation of the rest of the effective action
\beq
\kappa g^2 a \, F_{\mu\nu}F_{\rho\sigma}\epsilon^{\mu\nu\rho\sigma}.
\eeq{anom5}
The pseudoscalar $a$ is not canonically normalized, since its kinetic term is $(1/2)J''(C)\partial_\mu b \partial^\mu b$. When written in terms of the canonically normalized field $\tilde a= \sqrt{- J''(C)}a$, eq.~\rf{anom5} defines a UV cutoff scale, $\Lambda_{UV}= \sqrt{-J''(C)}/\kappa g^2$, which must be larger than the Hubble scale during inflation $H={l O}(g)$. 

The resulting constraint is actually weaker than that we obtained from the modification to the  D-term equation~\rf{anom4}; specifically, in the case of the $-R+\alpha R^2$ theory, 
\beq
\Lambda_{UV}= \frac{\sqrt{-J''(C)}}{\kappa g^2} \sim \frac{16\pi^2}{ |C| g^2}\gg g \
\implies \ |C| \ll \frac{16\pi^2}{g^3} \sim 16\pi^2 \, 10^{15}.
\eeq{anom6}
Even if the UV cutoff scale is the Planck scale, we still have a bound which is easy to satisfy, $|C| \ll \frac{16\pi^2}{g^2} \sim 16\pi^2 \, 10^{10}$
corresponds to $80 \ll 16\pi^2 \, 10^{10}$. Thus, anomalies and  anomaly-like corrections are irrelevant for our class of models.

 \subsection{$\Phi(C)=-Ce^C$ case}
 
 We will start our discussion with the case where $\Psi_n(C)= C^{n-2}$, dual to pure supergravity model  with higher curvature terms, $-{1\over 2} R+{\alpha \over 2} R^2 + \xi_n R^n$. 
 For this case  the analysis will be more detailed and specific.
 This is the supergravity scalar/vector model which is dual to pure super gravitational model with higher curvature terms. The bosonic part of this action is
  \be
{\cal L}^\xi  = 
-{1\over 4} F^2 + {1\over 2} D^2  +  g (1+C^{-1}) D +\xi_4 \, g^4  (F^4+ D^4) C^2+...
\label{scal}\ee
It is dual to
$
{\cal L}^\xi(R)= -{1\over 2} R +{1\over 18 g^2} R^2 + \xi_4 R^{4}
$.
On the pure gravity side  the reason for making a choice of $\xi$ to be equal or greater that   $10^{22}$ in Planck units,  taken in \cite{Farakos:2013cqa},  is questionable and requires justification which was not given in \cite{Farakos:2013cqa}. At $\xi_4=0$ the potential in this model is given by ${9\over 8}   g^2 (1+C^{-1})^2$ where $C=- e^{\sqrt{2/3} \vp}$ and $\vp$ is a canonically normalized scalar, the potential is flat at large $\vp$ and inflation takes place and agrees with the data under condition that $g^2\sim 10 ^{-10}$.
When $\xi_4\gtrsim 10^{25}$  it can destroy the flatness of the potential and inflationary cosmology, according to either the simple reasoning given in the Section 2, or using the analysis performed on the scalar side \rf{scal} in \cite{Farakos:2013cqa}. A more complete analysis will be given below.

The potential for the canonical $F_{\mu\nu}$ and $D$ has the form $V_0 \sim   D^2\sim 10^{-10}$. At the beginning of the last N e-foldings of inflation in this theory $-C= e^{\sqrt {2/3} \vp} \sim 4N/3$, which is equal to $80$ for $N = 60$ \cite{Kallosh:2013hoa}. We will concentrate on the case $N = 60$, and return to more general $N$ later.

Ignoring factors $O(1)$, we find that the first correction to the potential $V_4$ and the ratio of this correction to $V_0$ are given by
\beq
V_4= \xi_4 \, g^{4} \, D^{4} C^{2} \sim \xi_4 \, 10^{-36} \qquad \Rightarrow \qquad {V_4\over V_0}\sim \xi_4 \, 10^{-26} \ .
\eeq{sm14a}
If we expect $\xi_4$ to be of order ${ O}(1)$, we see that the first correction is 26 orders of magnitude below the level when it would matter. It is instructive also to compute the corrections to the first and second derivative of the potential since they give contribution to the corrections to the observable slow-roll parameters and, consequently, to $1-n_{s}$ and $r$. Here we have to take into account that 
\be
V_0= { 9 \over 8} g^2 (1-  e^{-\sqrt{2/3}\vp})^2\, , \qquad V_0'\sim   g^2 e^{-\sqrt{2/3}\vp}
\ee
and therefore, for $D \sim g^{2}$, one has
\be
V_4'\sim \xi_4 \, g^{8} e^{2\sqrt{2/3}\vp}\, , \qquad {V_4'\over V_0'}\sim  \xi_4 \, g^{6} e^{3\sqrt{2/3}\vp}\sim  \xi_4\, 10^{-24}  \ .
\ee
This leads to a constraint of $\xi_{4}$ which is a bit stronger than the one obtained earlier, but it is still very weak:  $\xi_{4} \lesssim 10^{{24}}$. Comparison of corrections to $V''$  leads to an almost identical constraint.

Extending this analysis to an arbitrary number of e-foldings $N$ shows that
\be
{V_4'\over V_0'}\sim  \xi_4 \, g^{6} e^{3\sqrt{2/3}\vp}\sim  \xi_4\, 10^{-30} N^{3} \ .
\ee
This means that for $\xi_{4} = O(1)$ the term $V_{4}$ does not affect the last $10^{10}$ e-foldings of inflation. Moreover, an investigation similar to the one performed in \cite{Linde:1986fd} shows that the regime of eternal inflation  in this model starts at $e^{3\sqrt{2/3}\vp} \sim N \sim 10^{6}$, so the higher corrections for $\xi_{4} = O(1)$ do not affect the existence of a regime of eternal inflation in this scenario.

But what if the natural value of $\xi_{4}$ is actually very large? Here we will address this issue when working with the dual model  \rf{scal} which has the Higgsed $g\neq 0$ as well as the  de-Higgsed phase $g=0$, both well defined in  \cite{Ferrara:2013rsa}. A natural attempt to justify  the large number  $\xi_4\geq 10^{24}$ would be to make it inversely proportional to a small gauge coupling. There are 3 cases to consider.

Case 1. If we take $\xi_4=  {x\over g^6}$ as in \cite{Farakos:2013cqa}, the theory on scalar side has no limit to $g=0$, because the potential blows up in the limit:
\be
{\cal L}^\xi  = 
-{1\over 4} F^2 + {1\over 2} D^2  + g J'(C) D +{x \over  g^2}  (F^4+ D^4) C^2 \ ,
\ee
 all other terms in our action depend on $g$ in vertices. Thus we rule out taking $\xi=  {x\over g^6}$ since the  limit to the de-Higgsed phase blows up, 
and this not acceptable.
\

Case 2. $\xi_4= {y\over g^4}$
\be
{\cal L}^\xi  = 
-{1\over 4} F^2 + {1\over 2} D^2  + g J'(C) D +y    (F^4+ D^4) C^2 \ .
\ee
In the limit $g=0$ we find
\be
{\cal L}^\xi  = 
-{1\over 4} F^2 + {1\over 2} D^2   +y (F^4+ D^4) C^2 \ .
\ee
The solution for $D$ now is $D=0$, there are no corrections to the potential  at $g=0$. However, for $y\neq 0$  there is an interaction between Maxwell curvature and scalars at $g=0$, which reminds of a BI interaction
\be
{\cal L}^\xi  = 
-{1\over 4} F^2   +y F^4 e^{2\phi} \ .
\ee
A closer inspection shows that it is not possible to associate the new coupling $y$ with  a Born-Infeld coupling \cite{Cecotti:1986gb}, \cite{Aschieri:2008ns}. In our case, at $g=0$, the Maxwell term $-{1\over 4} F^2$ is decoupled from scalars, therefore 
 we  conclude that  $y$ must be zero since it does not lead to a consistent supersymmetric BI model in the de-Higgsed phase. 
 
An alternative argument can be given, leading to an analogous conclusion. The one-loop UV divergences in N=1 supergravity interacting with matter are proportional to terms with a square of the energy-momentum tensor, $T_{\mu\nu}^2$. In the de-Higgsed phase the vectors are decoupled from scalars, their energy-momentum depends only on a Maxwell curvature squared, so the term $F^4 e^{2\phi}$ is not part of the $T_{\mu\nu}^2\sim F^4+..$ term, which also suggests that $y=0$.

Case 3. $\xi_4= {z\over g^2}$
\be
{\cal L}^\xi  = 
-{1\over 4} F^2 + {1\over 2} D^2  + g J'(C) D +x  g^2  (F^4+ D^4) C^2 \ .
\ee
With $\xi= z\times 10^{10}$ one still has to explain where the other factor $10^{14}$ is coming from. Therefore, even if this case is possible, it is not motivated and not indicating that the flatness of the potential is in any danger.  We therefore skip this case and proceed with the analysis for $\xi_4={ O} (1)$ since there is no reason to assume any other value. In this case, the first correction is negligible during the last 60 e-foldings of inflation.

The higher order terms on pure gravity side are $\xi_n R^n$ and on scalar/vector side are given by 
\beq
{\cal L}_{n} = \xi_{n} \, g^{n} \, D^{n} C^{n-2}  \ .
\eeq{sm13b}
Numerically during the last 60 e-foldings of inflation $C\sim 80$ whereas $gD\sim 10^{-10}$. Therefore all these terms, for $\xi_n$ independent on $g$, become
 \beq
{\cal L}_{n} \leq  \xi_{n} \, 10^{-8n-4}  \ .
\eeq{sm13c}
All these terms are incredibly small for $\xi_n$ independent on inverse powers of $g^2$. And since there is no known reason to claim such a dependence and in the simple case of $\xi_4$ we argued against it, we believe that the available arguments about higher order corrections presenting a danger to inflation are not justified.

\subsection{Models with generic $\Phi(C)$}

The form of higher order corrections we found in these models at every higher level has some arbitrary functions of the scalar field $C$ as shown in eq. \rf{caseN}. We start with the first correction, the potential is
\be
V (C) =  {1\over 2} \hat D^2  + g J'(C) \hat D +  \xi_{4}   \, g^4 \Psi_{4} (C) \hat D^{4} \ .
\label{case4}\ee
where $\hat D(C)$ solves the following equation
\be
- g J'(C) = \hat D(C) + 4 \xi_{4}   \, g^4 \Psi_{4} (C) \hat D^3(C) \ ,
\ee
and $\Psi_{4} (C)$ is some function. In the example considered in the previous section, the function was $\Psi_{4} (C)=C^2$. In our general class of models with $\Phi(C)= e^{{2\over 3} J(C)}$ the function $\Psi_{4} (C)$ is not known. However, observational data are available only for a part of the last $60$ e-foldings of inflation, which translates into a very limited range of the variable $C$. For example, in the theory ${g^{2}\over 2}\vp^{2}$, where $\vp = C$ \cite{Ferrara:2013rsa}, the most important range of $C$ is $C= O(10)$. Therefore in our analysis we will treat $\xi_{4}  \Psi_{4} (C)$ as an unknown constant, and place constraints on its value at the time corresponding to $N \sim 60$.

The exact solution of equation for $\hat D(C)$ and the exact potential are known in this case, see \cite{cfgvp}. However, in view 
 of the smallness of $g$  it  is simpler to compare the value of  $\Delta V$ with $V_0$, as we did above. As a particular example, one may consider a theory  of ${g^{2}\over 2}\vp^{2}$ type, where $g^{2} \sim 10^{{-10}}$,  $D^{2} \sim 10^{-10}$ and $C = O(10)$ during the last $60$ e-foldings of inflation. We find
\beq
V_4= \xi_4 \, g^{4} \, D^{4} \Psi_{4} (C) \sim \xi_4 \, 10^{-30}   \Psi_{4} (C)\,  V_0\qquad \Rightarrow \qquad {V_4\over V_0}\sim  10^{-30}  \,  \xi_4 \Psi_{4} (C) \ .
\eeq{sm14b}
Thus, we find a constraint 
\be
\xi_4 \Psi_{4} (C)\lesssim 10^{30} \ .
\ee
Unless one knows why the term $V_4$ should contain a huge coefficient $\xi_4 \Psi_{4} (C)\geq 10^{30}$, there is no reason to expect that the term $V_4$ may affect the dynamics of the last 60 e-foldings of inflation.

For  higher order terms we need to look at
\be
V_n= \xi_n \, g^{n} \, D^{n} \Psi_{n} (C) \sim \xi_n 10^{-10 (n-1)}\,  \Psi_n(C) \, V_0 \qquad \Rightarrow \qquad {V_n\over V_0}\sim  10^{-10 (n-1)}  \,  \xi_n \Psi_{4} (C) 
\ee
and the constraint is
\be
\xi_n \Psi_{n} (C) < 10^{10 (n-1)} \ .
\ee
Here we see again the higher order corrections become significant and comparable with the uncorrected inflationary potential $V_0$ only if the coefficients $\xi_n \Psi_{n} (C)$ are enormously large. As long as the higher derivative corrections do not come with these huge coefficients in front of them, they do not affect inflation, if the proper choice of the function $J(C)$ defining $V_0= {g^2\over 2} J'(C)^2$ was made to provide the required flatness of the potential in our minimal supergravity class of models \cite{Ferrara:2013rsa}.

\subsection{New minimal supergravity, massive  vector multiplets and the cosmological constant}
Now we will discuss a possibility of adding to our models a supersymmetric cosmological constant term. In the old minimal supergravity framework one could have started with the action
\beq
{\cal L}= -S_0 \bar{S}_0 \Phi(\Lambda+\bar{\Lambda} + g V)_D + {1\over 2 } [W_\alpha(V) W^\alpha(V) + h.c]_F  + \lambda [S_0^3+ h.c]_F \ ,
\eeq{CC}
where $\lambda $ is a constant. This terms breaks a superconformal $\mathbf{ R}$-symmetry of the model and  it is not dual to the new minimal supergravity anymore.
The resulting potential\footnote{ Incidentally, \rf{potential} gives the most general shift-symmetric potential of a chiral multiplet charged with respect to a gauged shift symmetry. The model contained in \cite{Dall'Agata:2013ksa} is obtained for $J=\log C$, and in the case $g^2=2 \lambda^2$ the potential vanishes.} is
$V= V_F+ V_D$
\be
V= \lambda^2 e^{-2J(C)} \Big ( 2 (J'(C) )^2 |(J'')^{-1}| - 3 \Big )+ {g^2\over 2} (J'(C))^2 \ .
\label{potential}\ee
Here we take into account that $-2J(C)= K= -3 \log \Phi$. 
This potential may have different vacua, corresponding to solutions of equation
\be
V'= J'\Big [ g^2 J'' + 2 \lambda^2 e^{-2J} (1+ 2 {J' \over J''} + {J' J'''\over (J'')^2})\Big ] = 0 \ .
\label{V'}\ee
For very small $\lambda$, the  stable Minkowski minimum, which was present at $\lambda=0$, becomes an AdS minimum, 
\be
V|_{J'=0}= -3\lambda^2 e^{-2J(C_0)}\,  \qquad J'(C_0)=0 \ .
\ee
However, 
as soon as  $\lambda^2 $ exceeds  ${g^2\over 2}   |J''(C) | e^{2J(C)} $ the former minimum becomes an AdS 
maximum, 
since
\be
V''|_{J'=0} =  g^2 (J'')^2  - |J'' | 2 \lambda^2 e^{-2J}  \ .
\ee
For example, in the theory with $J(C) = {3\over 2} (\log (-C) +C)$ the $\lambda=0$ Minkowski minimum becomes and AdS maximum for $\lambda^2 > {3\over 4} g^2 e^3$; at smaller $\lambda$ it remains an AdS minimum.

The second vacuum, if available, requires that the expression in square brackets in \rf{V'} vanishes. This case requires a separate study. It would be interesting to see whether one can find a new  Minkowski/dS vacuum due to vanishing of the expression in square brackets in \rf{V'}. These non-generic vacua are hard to find and it is not known whether they are available.

Therefore, generically the model (\ref{CC}) with $\lambda  \not = 0$ describes a collapsing Friedmann universe due to a negative cosmological constant. From the point of view of cosmology, the inevitable collapse of the universe is an undesirable feature of the model. It is therefore interesting that the new minimal supergravity formulation forbids such terms \cite{Ferrara:1983dh,Ferrara:1988qxa}. In old minimal supergravity we have to add a requirement of the superconformal $\mathbf{ R}$-symmetry to avoid a term $ \lambda [S_0^3+ h.c]_F $ and to preserve the  duality to  new minimal supergravity which exists only at $\lambda=0$.

\section{Summary}
 
In our investigation of the higher order corrections, we concentrated on the new-minimal supergravity with a massive vector multiplet and a  single real scalar identified with the inflaton field  \cite{Ferrara:2013rsa,vp}. This is a very economical and simultaneously very flexible framework, which allows to describe nearly arbitrary inflaton potentials. In comparison, the models based on chiral multiplets, which are capable of describing the inflaton field with an arbitrary potential, involve at least 4 scalar fields: the complex inflaton field and the complex sgoldstino, see e.g. \cite{Kallosh:2010ug}. However, during the long history of investigations of models with chiral multiplets we learned many ways to stabilize extra scalar fields and unify inflationary models with particles physics, which usually requires introduction of additional scalar fields anyway. The possibility to achieve this in the models with a massive vector multiplet is less explored and should be a subject of a separate investigation; see in this respect \cite{Ferrara:1983dh,Ferrara:1988qxa}. In this paper we limited ourselves to investigating the minimal  single-scalar supergravity  class of models  \cite{Ferrara:2013rsa,vp} and some of its features, including higher order supersymmetric corrections and their effect on the dynamics of inflation.   
 
Several new formal results were obtained in this paper; in particular, we found an explicit dual form of the supergravity model with $-{1\over 2} R +{1\over 18 g^2} R^2 +\sum_{n=4} \xi_n R^4$ bosonic action.
  A supersymmetry-inspired  expression for the bosonic, non-supersymmetric model of the form $-{1\over 2} R+ f(R) $ with arbitrary function $f(R)= {1\over 18 g^2} R^2 + \tilde f(R)$  can be given in the following  dual form 
  \beq
{\cal L}= -{1\over 2} R -  {1\over 2}   (\partial_\mu  \vp)^2  -  {D\over 2}    \Big (1-e^{-\sqrt{2/3}\vp}\Big ) + 
e^{-2\sqrt{2/3}\vp}    f  ( -D e^{\sqrt{2/3}\vp}   ) \ .
\eeq{smS}
Here gravity is of the Einstein form and there is a canonical scalar field $\vp$ and  a potential $V(\vp)$ which can be obtained in an explicit form when the auxiliary field $D$ is eliminated using its algebraic equations of motion. The auxiliary field $D$ in this bosonic theory in the supersymmetric case is a standard auxiliary $D$-field of the Maxwell multiplet, which is related to a $D$-term potential. Therefore the supergravity realization of the model with the bosonic term $f(R)= {1\over 18 g^2} R^2 + \sum_{n=4} \xi_n R^4$ has a dual form with a massive vector and a propagating scalar, that has the potential given in \rf{smS}. 

A manifestly superconformal action \rf{sm9}, with the bosonic part given by either the $-{1\over 2} R +{1\over 18 g^2} R^2 +\sum_{n=4} \xi_n R^4$ model or by the dual Einstein theory with a scalar and a massive vector, was given in this paper.  The gravity-scalar part of this superconformal action coincides with \rf{smS}. These models form a subset of minimal supergravity models of inflation studied in \cite{Ferrara:2013rsa,vp}, they are  defined by the Jordan function $\Phi(C)= -Ce^C$.

We have also generalized a set of supergravity models with a scalar and a massive vector discussed in \cite{Ferrara:2013rsa,vp} with a generic Jordan frame function $\Phi(C)$ and presented the superconformal higher superfield actions for these models in eq. \rf{corr}. 
We have used these formal results to study the effect of the higher power superfield actions on cosmological evolution.
The issue of higher corrections in supergravity models of inflation in general is complicated and model-dependent.
The class of models in  \cite{Ferrara:2013rsa,vp} allows us to describe a nearly unlimited class of inflationary potentials, which, however, have one common property: The potentials disappear in the limit $g\to 0$, where $g$ is the gauge coupling constant. As a result, one can consider models where the smallness of the amplitude of density perturbations produced during inflation is suppressed by the smallness of the constant $g^{2}$, which simultaneously suppresses higher order corrections. We studied these models and formulated the conditions which make higher order corrections vanishingly small during the last 60 e-foldings of inflation.

Some of our results apply to generic versions of the theory \cite{Ferrara:2013rsa}. However, we were able to obtain more specific results in the models of the type of $-R+\alpha R^{2}$, because we studied both the general supersymmetric version of the   $-R+\alpha R^{2} + \sum_{n=4} \xi_n R^n$ theory, which we constructed, and its dual version, where the gravity part is represented by the usual Einstein action and there is a scalar and a massive vector. This investigation, combined with an analysis of the possible dependence of higher order terms on $g^{2}$, suggests that higher order corrections to the inflationary model based on the supersymmetric generalization of the model $-R+\alpha R^{2}$ in the context of the theory developed in  \cite{Ferrara:2013rsa,cfps} do not affect the inflationary regime in these theories.

\subsection*{Acknowledgements}
We are grateful to S. Cecotti and A. Van Proeyen for stimulating discussions.
 The work of RK and AL is supported by the SITP and by  
NSF Grant No. 0756174.   The work of RK is also supported by John Templeton Foundation. S.F.   is supported by ERC Advanced Investigator Grant n. 226455 {\em Supersymmetry, Quantum Gravity and Gauge Fields (Superfields)}.
M.P. is supported in part by NSF grant PHY-0758032. M.P. would like to thank CERN for its kind hospitality and the ERC Advanced Investigator Grant n. 226455 for support while at CERN.


\begin{thebibliography}{99}

%\cite{Ferrara:2013rsa}
\bibitem{Ferrara:2013rsa} 
  S.~Ferrara, R.~Kallosh, A.~Linde and M.~Porrati,
  ``Minimal Supergravity Models of Inflation,''
  arXiv:1307.7696 [hep-th].
  %%CITATION = ARXIV:1307.7696;%%
  %1 citations counted in INSPIRE as of 05 Aug 2013

\bibitem{vp} 
  A.~Van Proeyen,
``Massive Vector Multiplets In Supergravity,''
  Nucl.\ Phys.\ B {\bf 162}, 376 (1980).
  %%CITATION = NUPHA,B162,376;%%
  %7 citations counted in INSPIRE as of 18 Jul 2013
 


%\cite{Linde:1983gd}
\bibitem{Linde:1983gd} 
  A.~D.~Linde,
``Chaotic Inflation,''
  Phys.\ Lett.\ B {\bf 129}, 177 (1983).
  %%CITATION = PHLTA,B129,177;%%
  %1495 citations counted in INSPIRE as of 09 May 2013
  
    
   
   
\bibitem{Starobinsky:1980te} 
  A.~A.~Starobinsky,
``A New Type of Isotropic Cosmological Models Without Singularity,''
  Phys.\ Lett.\ B {\bf 91}, 99 (1980).
  %%CITATION = PHLTA,B91,99;%%
  %1537 citations counted in INSPIRE as of 28 May 2013 
  %\cite{Mukhanov:1981xt}
%\bibitem{Mukhanov:1981xt} 
  V.~F.~Mukhanov and G.~V.~Chibisov,
``Quantum Fluctuation and Nonsingular Universe. (In Russian),''
  JETP Lett.\  {\bf 33}, 532 (1981)
  [Pisma Zh.\ Eksp.\ Teor.\ Fiz.\  {\bf 33}, 549 (1981)].
  %%CITATION = JTPLA,33,532;%%
      %\cite{Starobinsky:1983zz}
%\bibitem{Starobinsky:1983zz} 
  A.~A.~Starobinsky,
``The Perturbation Spectrum Evolving from a Nonsingular Initially De-Sitter Cosmology and the Microwave Background Anisotropy,''
  Sov.\ Astron.\ Lett.\  {\bf 9}, 302 (1983).
  %%CITATION = SALED,9,302;%%
   
  
%\cite{Kofman:1985aw}
\bibitem{Kofman:1985aw} 
  L.~A.~Kofman, A.~D.~Linde and A.~A.~Starobinsky,
``Inflationary Universe Generated by the Combined Action of a Scalar Field and Gravitational Vacuum Polarization,''
  Phys.\ Lett.\ B {\bf 157}, 361 (1985).
  %%CITATION = PHLTA,B157,361;%%
  %\cite{Kofman:1986yt}
  
      %\cite{Whitt:1984pd}
\bibitem{Whitt:1984pd} 
  B.~Whitt,
``Fourth Order Gravity as General Relativity Plus Matter,''
  Phys.\ Lett.\ B {\bf 145}, 176 (1984).


  %\cite{Cecotti:1987sa}
\bibitem{Cecotti:1987sa} 
  S.~Cecotti,
  ``Higher Derivative Supergravity Is Equivalent To Standard Supergravity Coupled To Matter. 1.,''
  Phys.\ Lett.\ B {\bf 190}, 86 (1987).
  %%CITATION = PHLTA,B190,86;%%
  %21 citations counted in INSPIRE as of 09 Jul 2013
  
  

  
  
   %\cite{Cecotti:1987qe}
\bibitem{cfps} 
  S.~Cecotti, S.~Ferrara, M.~Porrati and S.~Sabharwal,
``New Minimal Higher Derivative Supergravity Coupled To Matter,''
  Nucl.\ Phys.\ B {\bf 306}, 160 (1988).
  %%CITATION = NUPHA,B306,160;%%
  %12 citations counted in INSPIRE as of 18 Jul 2013
  


    %\cite{Farakos:2013cqa}
\bibitem{Farakos:2013cqa} 
  F.~Farakos, A.~Kehagias and A.~Riotto,
 ``On the Starobinsky Model of Inflation from Supergravity,''
  arXiv:1307.1137 [hep-th].
  %%CITATION = ARXIV:1307.1137;%%
  %1 citations counted in INSPIRE as of 15 Jul 2013
  
    %\cite{Ellis:2013nxa}
\bibitem{Ellis:2013nxa} 
  J.~Ellis, D.~V.~Nanopoulos and K.~A.~Olive,
 ``Starobinsky-like Inflationary Models as Avatars of No-Scale Supergravity,''
  arXiv:1307.3537 [hep-th].
  %%CITATION = ARXIV:1307.3537;%%
  %5 citations counted in INSPIRE as of 29 Aug 2013

 %\cite{Fre:2013dha}
\bibitem{Fre:2013dha} 
P.~Fr$\acute{\rm e}$, A.~Sagnotti and A.~S.~Sorin,
``Integrable Scalar Cosmologies I. Foundations and links with String Theory,''
  arXiv:1307.1910 [hep-th].
  %%CITATION = ARXIV:1307.1910;%%
  P.~Fr$\acute{\rm e}$ and A.~S.~Sorin,
  ``Inflation and Integrable one-field Cosmologies embedded in Rheonomic Supergravity,''
  arXiv:1308.2332 [hep-th].
  %%CITATION = ARXIV:1308.2332;%% 
  
      %\cite{Kallosh:2010ug}
\bibitem{Kallosh:2010ug} 
  R.~Kallosh, A.~Linde,
  ``New models of chaotic inflation in supergravity,''
  JCAP {\bf 1011}, 011 (2010)
  [arXiv:1008.3375 [hep-th]].
  %%CITATION = ARXIV:1008.3375;%%
  %44 citations counted in INSPIRE as of 31 Mar 2013
  %\cite{Kallosh:2010xz}
%\bibitem{Kallosh:2010xz} 
  R.~Kallosh, A.~Linde, T.~Rube,
  ``General inflaton potentials in supergravity,''
  Phys.\ Rev.\ D {\bf 83}, 043507 (2011)
  [arXiv:1011.5945 [hep-th]].
  %%CITATION = ARXIV:1011.5945;%%
  %23 citations counted in INSPIRE as of 31 Mar 2013
%\cite{Kallosh:2011qk}
%\bibitem{Kallosh:2011qk} 
  R.~Kallosh, A.~Linde, K.~A.~Olive and T.~Rube,
``Chaotic inflation and supersymmetry breaking,''
  Phys.\ Rev.\ D {\bf 84}, 083519 (2011)
  [arXiv:1106.6025 [hep-th]].
  %%CITATION = ARXIV:1106.6025;%%
  %18 citations counted in INSPIRE as of 13 Jun 2013

%\cite{Linde:2005ht}
\bibitem{Linde:2005ht} 
  A.~D.~Linde,
``Particle physics and inflationary cosmology,''
  Contemp.\ Concepts Phys.\  {\bf 5}, 1 (1990)
  [hep-th/0503203].

 
%\cite{Freese:1990rb}
\bibitem{Freese:1990rb} 
  K.~Freese, J.~A.~Frieman and A.~V.~Olinto,
``Natural inflation with pseudo-Nambu-Goldstone bosons,''
  Phys.\ Rev.\ Lett.\  {\bf 65}, 3233 (1990).
  %%CITATION = PRLTA,65,3233;%%
  %375 citations counted in INSPIRE as of 23 Aug 2013


%\cite{Silverstein:2008sg}
\bibitem{Silverstein:2008sg} 
  E.~Silverstein and A.~Westphal,
``Monodromy in the CMB: Gravity Waves and String Inflation,''
  Phys.\ Rev.\ D {\bf 78}, 106003 (2008)
  [arXiv:0803.3085 [hep-th]].
  %%CITATION = ARXIV:0803.3085;%%
L.~McAllister, E.~Silverstein and A.~Westphal,
``Gravity Waves and Linear Inflation from Axion Monodromy,''
  Phys.\ Rev.\ D {\bf 82}, 046003 (2010)
  [arXiv:0808.0706 [hep-th]].

  
  %\cite{Salopek:1988qh}
\bibitem{Salopek:1988qh}
  D.~S.~Salopek, J.~R.~Bond and J.~M.~Bardeen,
``Designing density fluctuation spectra in inflation,''
  Phys.\ Rev.\  \textbf{D40}, 1753 (1989).
  %%CITATION = PHRVA,D40,1753;%%

   \bibitem{Sha-1}  F.~L.~Bezrukov and M.~Shaposhnikov, ``The Standard
Model Higgs boson as the inflaton,'' Phys.\ Lett.\ B {\bf 659}, 703
(2008) [arXiv:0710.3755 [hep-th]].

    %\cite{Ferrara:2010in}
\bibitem{Ferrara:2010in} 
  S.~Ferrara, R.~Kallosh, A.~Linde, A.~Marrani, A.~Van Proeyen and ,
 ``Superconformal Symmetry, NMSSM, and Inflation,''
  Phys.\ Rev.\ D {\bf 83}, 025008 (2011)
  [arXiv:1008.2942 [hep-th]].


%\cite{Dong:2010in}
\bibitem{Dong:2010in} 
  X.~Dong, B.~Horn, E.~Silverstein and A.~Westphal,
``Simple exercises to flatten your potential,''
  Phys.\ Rev.\ D {\bf 84}, 026011 (2011)
  [arXiv:1011.4521 [hep-th]].
  %%CITATION = ARXIV:1011.4521;%%
  
  %\cite{Kallosh:2013hoa}
\bibitem{Kallosh:2013hoa} 
  R.~Kallosh and A.~Linde,
``Universality Class in Conformal Inflation,''
  JCAP {\bf 1307}, 002 (2013)
  [arXiv:1306.5220 [hep-th]].
  %%CITATION = ARXIV:1306.5220;%%
 R.~Kallosh and A.~Linde,
``Non-minimal Inflationary Attractors,''
  arXiv:1307.7938 [hep-th].
  %%CITATION = ARXIV:1307.7938;%%
  
  %\cite{Kallosh:2013pby}
\bibitem{Kallosh:2013pby} 
  R.~Kallosh and A.~Linde,
``Superconformal generalization of the chaotic inflation model $\frac{\lambda}{4} \phi^{4} - \frac{\xi}{2} \phi^{2}R$,''
  JCAP {\bf 1306}, 027 (2013)
  [arXiv:1306.3211 [hep-th]].
  %%CITATION = ARXIV:1306.3211;%%
  

 
   \bibitem{cfgvp} 
  E.~Cremmer, S.~Ferrara, L.~Girardello and A.~Van Proeyen,
  ``Yang-Mills Theories with Local Supersymmetry: Lagrangian, Transformation Laws and SuperHiggs Effect,''
  Nucl.\ Phys.\ B {\bf 212}, 413 (1983).
  %%CITATION = NUPHA,B212,413;%%
  %956 citations counted in INSPIRE as of 23 Jul 2013  
  

%\cite{Ferrara:1978rk}
\bibitem{Ferrara:1978rk} 
  S.~Ferrara, M.~T.~Grisaru and P.~van Nieuwenhuizen,
  ``Poincare And Conformal Supergravity Models With Closed Algebras,''
  Nucl.\ Phys.\ B {\bf 138}, 430 (1978).
  %%CITATION = NUPHA,B138,430;%%
  %71 citations counted in INSPIRE as of 15 Aug 2013
  
  
%\cite{Ketov:2009sq}
\bibitem{Ketov:2009sq} 
  S.~Ketov,
  ``F(R) supergravity,''
  AIP Conf.\ Proc.\  {\bf 1241}, 613 (2010)
  [\href{http://www.arXiv.org/abs/0910.1165}{{\tt 0910.1165}}]
  %%CITATION = ARXIV:0910.1165;%%
  %10 citations counted in INSPIRE as of 09 Jul 2013
%\cite{Ketov:2012yz}
%\bibitem{Ketov:2012yz} 
  S.~V.~Ketov,
  ``Supergravity and Early Universe: the Meeting Point of Cosmology and High-Energy Physics,''
  Int.\ J.\ Mod.\ Phys.\ A {\bf 28}, 1330021 (2013)
  [arXiv:1201.2239 [hep-th]].
  %%CITATION = ARXIV:1201.2239;%%
  %3 citations counted in INSPIRE as of 09 Jul 2013

  
%\cite{Kugo:1983mv}
\bibitem{Kugo:1983mv} 
  T.~Kugo and S.~Uehara,
  ``N=1 Superconformal Tensor Calculus: Multiplets With External Lorentz Indices And Spinor Derivative Operators,''
  Prog.\ Theor.\ Phys.\  {\bf 73}, 235 (1985).
  %%CITATION = PTPKA,73,235;%%
  %46 citations counted in INSPIRE as of 27 Jul 2013
  
  
  
   %\cite{Green:1984sg}
\bibitem{Green:1984sg} 
  M.~B.~Green and J.~H.~Schwarz,
  ``Anomaly Cancellation in Supersymmetric D=10 Gauge Theory and Superstring Theory,''
  Phys.\ Lett.\ B {\bf 149}, 117 (1984).
  %%CITATION = PHLTA,B149,117;%%
  %2165 citations counted in INSPIRE as of 29 Aug 2013
  
  

  
     
    \bibitem{nielsen}
N.~K.~Nielsen and H.~Romer,
  ``Nonabelian Anomaly For Spin 3/2,''
  Phys.\ Lett.\ B {\bf 154}, 141 (1985).


  
  %\cite{Cecotti:1986gb}
\bibitem{Cecotti:1986gb} 
  S.~Cecotti and S.~Ferrara,
  ``Supersymmetric Born-infeld Lagrangians,''
  Phys.\ Lett.\ B {\bf 187}, 335 (1987).
  %%CITATION = PHLTA,B187,335;%%
  %85 citations counted in INSPIRE as of 07 Aug 2013
  
 %\cite{Aschieri:2008ns}
\bibitem{Aschieri:2008ns} 
  P.~Aschieri, S.~Ferrara and B.~Zumino,
  ``Duality Rotations in Nonlinear Electrodynamics and in Extended Supergravity,''
  Riv.\ Nuovo Cim.\  {\bf 31}, 625 (2008)
  [arXiv:0807.4039 [hep-th]].
  %%CITATION = ARXIV:0807.4039;%%
  %35 citations counted in INSPIRE as of 28 Aug 2013 
  
  %\cite{Linde:1986fd}
\bibitem{Linde:1986fd} 
  A.~D.~Linde,
``Eternally Existing Self-reproducing Chaotic Inflationary Universe,''
  Phys.\ Lett.\ B {\bf 175}, 395 (1986).
  %%CITATION = PHLTA,B175,395;%%
  %477 citations counted in INSPIRE as of 01 Sep 2013
  
  %\cite{Dall'Agata:2013ksa}
\bibitem{Dall'Agata:2013ksa} 
  G.~Dall'Agata and F.~Zwirner,
``A new class of N=1 no-scale supergravity models,''
  arXiv:1308.5685 [hep-th].
  %%CITATION = ARXIV:1308.5685;%%



  
  %\cite{Ferrara:1983dh}
\bibitem{Ferrara:1983dh} 
  S.~Ferrara, L.~Girardello, T.~Kugo and A.~Van Proeyen,
  ``Relation Between Different Auxiliary Field Formulations Of N=1 Supergravity Coupled To Matter,''
  Nucl.\ Phys.\ B {\bf 223}, 191 (1983).
  %%CITATION = NUPHA,B223,191;%%
  %115 citations counted in INSPIRE as of 02 Sep 2013

  
  
  %\cite{Ferrara:1988qxa}
\bibitem{Ferrara:1988qxa} 
  S.~Ferrara and S.~Sabharwal,
``Structure Of New Minimal Supergravity,''
  Annals Phys.\  {\bf 189}, 318 (1989).
  %%CITATION = APNYA,189,318;%%

   


 \end{thebibliography}
\end{document}